\documentclass[fleqn,usenatbib]{mnras}

\usepackage{newtxtext,newtxmath}
 
\usepackage[T1]{fontenc}

\DeclareRobustCommand{\VAN}[3]{#2}
\let\VANthebibliography\thebibliography
\def\thebibliography{\DeclareRobustCommand{\VAN}[3]{##3}\VANthebibliography}

\usepackage{graphicx}	% Including figure files
\usepackage{amsmath}	% Advanced maths commands
\usepackage{amssymb}	% Extra maths symbols

\newcommand{\pdd}[2]{\frac{\partial #1}{\partial #2}}
\newcommand{\vsig}{v_\phi/\sigma_r}
\newcommand{\vmaxsig}{v_\mathrm{max}/\langle \sigma \rangle}

\title[On the origin of surprisingly cold gas discs]{On the origin of surprisingly cold gas discs in galaxies at high redshift}

\author[Kretschmer et al.]{
Michael Kretschmer,$^{1}$\thanks{E-mail: michael.kretschmer@physik.uzh.ch}
Avishai Dekel,$^{2}$
Romain Teyssier$^{1,3}$
\\
$^{1}$Institute for Computational Science, University of Zurich, Winterthurerstrasse 190, CH-8057 Zurich, Switzerland\\
$^{2}$Center for Astrophysics and Planetary Science, Racah Institute of Physics, The Hebrew University, Jerusalem 91904, Israel\\
$^{3}$Department of Astrophysical Sciences, Princeton University, 4 Ivy Lane, Princeton, New Jersey, 08544, United States
}

\date{Accepted XXX. Received YYY; in original form ZZZ}

\pubyear{2021}

\begin{document}
\label{firstpage}
\pagerange{\pageref{firstpage}--\pageref{lastpage}}
\maketitle

\begin{abstract}
We address the puzzling observational indications for very ``cold'' galactic discs at redshifts $z \gtrsim 3$, an epoch when discs are expected to be highly perturbed. Using a high-resolution cosmological zoom-in simulation, we identify such a cold disc at $z\sim 3.5$, with a rotation velocity to velocity dispersion ratio of $\vsig \simeq 5$ for the total gas. It forms as a result of a period of intense accretion of co-planar, co-rotating gas via cold cosmic-web streams.
This thin disc survives for $\sim 5$ orbital periods, after which it is disrupted by mergers and counter-rotating streams, longer but consistent with our estimate that a galaxy of this mass ($M_\star\sim10^{10}\mathrm{M_\odot}$) typically survives merger-driven spin flips for $\sim 2-3$ orbital periods.
We find that $\vsig$ is highly sensitive to the tracer used to perform the kinematic analysis. While it is $\vsig \simeq 3.5$ for atomic HI gas, it is $\vsig \simeq 8$ for molecular CO and H$_2$. This reflects the confinement of molecular gas to cold, dense clouds that reside near the disc mid-plane, while the atomic gas is spread into a turbulent and more extended thicker disc.
The proposed mechanisms is a theoretical proposal that has not been validated yet with proper statistical measurements and it remains unclear whether it occurs frequently enough to explain the multiple discoveries of cold gas disks in high-z galaxies.
\end{abstract}


%%%%%%%%%%%%%%%%%%%%%%%%%%%%%%%%%%%%%%%%%%%%%%%%%%

%%%%%%%%%%%%%%%%% BODY OF PAPER %%%%%%%%%%%%%%%%%%
\section{Introduction}
In our current picture of galaxy formation, razor thin gas discs only form at low redshift ($z < 1.5$), while high-redshift galaxies are predicted to be highly perturbed through various dynamical instabilities, driven by intense stream-fed accretion including frequent mergers and counter-rotating gas streams \citep[][]{2007ApJ...670..237B,2008ApJ...687...59G,2009ApJ...703..785D,2011ApJ...733L..11R,2012MNRAS.423.3616D,2013ApJ...768..164A,2014MNRAS.438.1870D,2015MNRAS.450.2327Z,2017MNRAS.465.2643C,2019MNRAS.490.3196P,2020MNRAS.496.5372D}.
Major mergers also produce disc spin flips that occur on a time-scale shorter than the orbital period in galaxies that reside in low mass halos with virial masses $M_\mathrm{vir}<2 \times 10^{11}\mathrm M_\odot$ \citep[][]{2020MNRAS.493.4126D}.

Additionally, efficient accretion of filamentary dense and cold gas constantly brings new material to the disc, causing high gas densities which enable large star formation rates (SFR) and intense associated supernova (SN) feedback \citep[][]{2005MNRAS.363....2K,2009MNRAS.395..160K,2006MNRAS.368....2D,2008MNRAS.390.1326O,2009Natur.457..451D,2009ApJ...691.1168H,2017MNRAS.465.1682H,2021arXiv210611981T}.
Cosmic gas accretion has been carefully studied in cosmological simulations \citep[][]{2012MNRAS.422.1732D,2015MNRAS.449.2087D,2013ApJ...769...74S,2021arXiv211005384C} and similar results are being indicated in observations \citep[e.g.][]{2016ApJ...820..121B,2019NatAs...3..822M}. While the accreting angular momentum outside and at the virial radius may be tilted relative to the central galaxy spin, it is becoming more aligned with the disc spin as the gas penetrates the inner halo.

At high redshift, SNe explosions are indeed highly clustered and inject large amounts of energy in the surrounding interstellar medium (ISM), launching powerful outflows and preventing the formation of thin discs \citep[][]{2017ApJ...834...25K, 2019MNRAS.483.3647G,2018MNRAS.481.3325F,2020MNRAS.492...79M}.

On the other hand, these cold streams can also enable the growth of extended discs if the accreted material is mostly co-rotating and co-planar for a sufficiently long time  \citep[][]{2012MNRAS.423.1544S,2012MNRAS.422.1732D,2015MNRAS.449.2087D,2013ApJ...769...74S,2020MNRAS.497.4346K}.
These large discs survive over a long time-scale only when merger events are rare enough. It is only under sufficiently quiescent conditions that galaxies are able to develop thin and cold discs with a rotation velocity significantly larger than the velocity dispersion \citep[][]{2016AJ....152..157L}.
As shown by \cite{2020MNRAS.493.4126D}, moderately cold gas discs (with a rotation velocity $V$ to velocity dispersion $\sigma$ ratio of $V/\sigma \sim 4$) that can survive for a few orbital times are indeed expected in large enough halos, above a critical mass of $M_\mathrm{vir}\sim 2 \times 10^{11}\mathrm M_\odot$, and at all redshifts.
This mass threshold corresponds to a number density of discs that increases with time.
Although we do expect more rotation-supported discs at low redshift, this simple picture also allows for the formation of rotation-supported discs at high redshift in massive enough galaxies.

In fact, many observations have shown perturbed discs with $V/\sigma \gtrsim 3$ observed at $z\sim2$ \citep[][]{2017ApJ...843...46S,2020ApJ...902...98G}.
Recent observations however suggest the existence of massive $(M_\star>10^{10}\mathrm{M_\odot})$ rotation-dominated discs at high redshift $(z>4)$ with a surprisingly large $V/\sigma \sim 10$ \citep{2018MNRAS.479.5440L,2019MNRAS.487.4305S,2020Natur.584..201R,2021A&A...647A.194F,2020Natur.581..269N,2021Sci...371..713L,2021MNRAS.507.3952R,2021Sci...372.1201T}.
Actually, up to date, all high-z galaxies observed with ALMA at high enough resolution show cold gas disks, for a total number of 13 objects at $z>2.5$. Therefore, it is becoming evident that cold gas disks at high redshift cannot be an extremely rare phenomenon, posing a potential challenge to our understanding of galaxy formation.
This motivates our present paper, whose goal is to find and study a striking example of a particularly cold disc in a high-resolution cosmological simulation.

To study the gas kinematics in galaxies near and far, spatially resolved line emission is analysed through a technique known as integral field spectroscopy (IFS) \citep[][]{2020ApJ...902...98G,2018ApJS..238...21F}.
Typically, a Gaussian line profile is fitted to position-position-velocity data cubes resulting in detailed maps of the rotational velocity, as well as the velocity dispersion.

The choice of the emission line used for the observation is crucial, as it provides a tracer to a specific gas component.
Optical emission lines as H$\alpha$ and doubly ionised oxygen (OIII) have been routinely used at high redshift \citep[e.g.][]{2009ApJ...706.1364F, 2011A&A...528A..88G}.
It should be noted, that the relatively low spatial resolution, can result in beam smearing effects which may bias measurements to low $\vsig$ ratios \citep[][]{2016A&A...594A..77D}.
Another interesting line is emitted by single ionised carbon (CII) which tends to reside in the outskirts of dense molecular clouds \citep[][]{2018MNRAS.481.1976Z,2019MNRAS.486.4622C,2020A&A...643A.141M}.
Interestingly, CII which is used in ALMA observations, provides a better spatial resolution at $z=4-5$ than the before mentioned observations using H$\alpha$ and OIII at lower redshifts.

A common tracer for the hydrogen (H$_2$) molecular gas is for example the carbon-monoxide (CO). These molecules form in very dense environments, mostly encapsulated deep inside molecular clouds.
Also emission from neutral carbon (CI) has been used to trace molecular gas \citep[][]{2018MNRAS.479.5440L}.
Finally, the 21 cm line is also used to trace neutral atomic hydrogen (HI), for example with the Very Large Array (VLA) or in the future with the Square Kilometre Array (SKA), to perform an accurate kinematic analysis of the gas discs on larger scales \citep{1997MNRAS.290..533D,2013ApJ...765..136S,2015A&ARv..24....1G,2017MNRAS.466.4159I,2016AJ....152..157L,2019MNRAS.484.3267L,2019A&A...629A..59P,2019MNRAS.484.2234H}.

In this paper, we are interested in studying the kinematic properties of a simulated massive galaxy at high redshift. We address two main questions:
first, can we identify an epoch for which our simulated gas disc appears as strongly rotation-dominated? Second, can different emission lines trace different gas components with different kinematic properties?
To investigate these questions, we have structured our paper as follows:
in \autoref{section:methods} we present our simulation methodology, together with an overview of our adopted galaxy formation model. 
We present our results in \autoref{section:results}, focusing on the kinematic properties of our simulated galaxies and on the impact of the gas tracer.
Finally, we discuss the implications of our work in \autoref{section:discussion} and conclude in \autoref{section:conclusions}.

\section{Methods}
\label{section:methods}

We analyse a zoom-in simulation of a galaxy that was performed using the adaptive mesh refinement (AMR) code \texttt{RAMSES} \citep{2002A&A...385..337T}.
The simulation methods are described in detail in \cite{2020MNRAS.492.1385K} and \cite{2021MNRAS.503.5238K,2020MNRAS.497.4346K}, but we briefly summarise their main characteristics in the following sections.

\subsection{Halo selection and initial conditions}

We have used a parent $N$-body simulation with $512^3$ dark matter (DM) particles in a periodic box of size 25~$h^{-1}$~Mpc. We have selected several halos at $z=0$ with virial masses in the range $M_\mathrm{vir}=(0.75 - 1.5) \times 10^{12}\mathrm{M_\odot}$. The virial mass was calculated using a spherical over-density $\Delta(z)=(18\pi^2-82\Omega_\Lambda(z)-39\Omega_\Lambda(z)^2)/\Omega_m(z)$, relative to the mean mass density, where $\Omega_\Lambda$ and $\Omega_m(z)$ are the dark energy density parameter and the matter density parameter \citep{1998ApJ...495...80B}.
We have also required the halos to be without any major merger (with mass-ratios smaller than 3:1) after $z=1$ and to be sufficiently in isolation at $z=0$ such that there exists no halo more massive than half $M_\mathrm{vir}$ within 5 virial radii. This relatively arbitrary selection process led to a sample of nine halos. Our findings presented later, are not biased by the selection criteria.
We have then performed zoom-in cosmological simulations of these nine halos with a state-of-the-art galaxy formation model, resulting in a small catalogue of nine galaxies that we called the MIGA (MIchael's GAlaxies) catalogue (see below).

The galaxy that we analyse in this paper is a higher-resolution run of one particular MIGA galaxy which was also analysed in \cite{2020MNRAS.497.4346K} and ended up as a large, bulge-dominated disc at $z=0$.
We chose this particular galaxy because it also features an extended gas disc at high redshift, which is precisely the topic of this paper.

To make this higher-resolution simulation computationally feasible, we have created a new, smaller zoom-in Lagrangian region. For this, we extracted the DM particles inside $1.3 R_\mathrm{vir}$ at scalefactor $a_\mathrm{exp}=0.34$ ($z=1.94$) from the original MIGA simulation and traced them back to their positions at $z=100$, defining that way a new smaller zoom-in volume.
The smallest cells in the simulation have sizes $\Delta x_\mathrm{min}=27.5$~pc. The resolution was kept roughly constant in physical units by progressively releasing new refinement levels. The mass of the DM particles is $m_\mathrm{DM}=2.5 \times 10^{4}$M$_\odot$ and the initial baryonic mass is $m_\mathrm{bar}=3.6 \times 10^{3}$M$_\odot$.

\subsection{The MIGA catalogue}

The MIGA catalogue (MIchael's GAlaxies) is a simulation suite consisting of nine galaxies evolved to $z=0$ with a resolution of $55$~pc in the smallest cells. The halos at $z=0$ have masses similar to the Milky Way (MW) halo, but they host galaxies with very diverse morphologies. The stellar systems of four galaxies are disc-dominated, with large disc-to-total ratios $D/T>0.5$, where we defined the disc to be the sum of the stars that have eccentricities $\epsilon>0.5$ \citep[see, e.g.][]{2003ApJ...597...21A,2018MNRAS.473.1930E,2018MNRAS.477.4915O,2019ApJ...883...25P}. The other five galaxies are bulge-dominated, with $0.2<D/T<0.5$.

The final stellar masses are in excellent agreement with the values predicted using the abundance matching technique \citep{2013ApJ...770...57B}. In fact, all the MIGA galaxies scatter around the predicted relation, within the $1\sigma$ confidence interval, except for one case. This case corresponds to our most massive stellar system, likely lacking the feedback of an active galactic nucleus (AGN).

The galaxies all feature episodic starburst events, with large star formation efficiencies around 10\% or more, mostly triggered by major mergers at high redshift. They all end up as quiescent discs at $z=0$ with SFR$\sim 1 \mathrm{M_\odot yr^{-1}}$ and rather small star formation efficiencies around $1\%$ or less. 
Our efficient SN feedback is able to
remove large amounts of gas from the halo at high redshift, such that the final baryon fraction inside $R_\mathrm{vir}$ is typically $f_\mathrm{bar} = (M_\star + M_\mathrm{gas})/(\Omega_b/\Omega_m \, M_\mathrm{DM}) \simeq 0.4$. At the same time, a relatively large gas fraction of $f_\mathrm{gas}=M_\mathrm{gas}/(M_\star + M_\mathrm{gas})\simeq 20\%$ is present inside the disc between $z=0$ and $z=2$. 
These results depend strongly on our adopted recipe for star formation and feedback that we summarise below.

\subsection{The need for a subgrid model for molecular clouds}

Current cosmological galaxy formation simulations reach a spatial resolution between tens to hundreds of parsecs, barely resolving the largest molecular clouds in the Galaxy \citep[][]{2014MNRAS.442.1545C,2017MNRAS.472.2356M,2017MNRAS.467..179G,2018MNRAS.480..800H,2019MNRAS.490.4447W,2020MNRAS.491.1656A,2020MNRAS.491.3461B}.
While isolated galaxy simulations reach resolutions that allow the study of clouds \citep[][]{2016MNRAS.458.3528H,2017MNRAS.471.2151H,2019ApJ...879L..18L,2020ApJ...891....2L,2019MNRAS.482.1304E,2020ApJ...890..155E,2021MNRAS.501.5597G}, cosmological galaxy formation simulations in general are not capable of doing so. 

We are particularly interested in molecular clouds, since they form stars and host molecular tracers.
This led us to design a model that enables us to recover the properties of molecular clouds below the resolution limit \citep[for similar models see, e.g.][]{2009ApJ...693..216K,2009ApJ...707..954P,2011MNRAS.418..664N,2012MNRAS.426.2142L,2012MNRAS.425.3058C,2016MNRAS.461...93P,2018MNRAS.473..271V}.

In the next section, we describe the main components of our subgrid model, starting with our model to compute the velocity dispersion of the subgrid (or micro-) turbulence. We then describe how we use this information to recover the probability distribution function (PDF) of the gas cloud densities inside each cell and finally how we can obtain molecular fractions and star formation efficiencies from this PDF.

\subsection{Subgrid model for turbulence}

We model unresolved turbulent flows using a sub-grid scale (SGS) model to describe turbulent effects at the macroscopic scale \citep[][]{2006A&A...450..265S,2011A&A...528A.106S}. We introduce a new fluid variable for the turbulent kinetic energy
$K_T= 3/2 \rho \sigma_T^2$,
with the turbulent (one-dimensional) velocity dispersion $\sigma_T$.
We follow the standard formalism of implicit large eddy simulation (implicit LES) \citep[][]{1963MWRv...91...99S,2006A&A...450..265S,2015LRCA....1....2S,2019PhRvE.100d3116S} and introduce an extra equation \citep[][]{2011A&A...528A.106S,2014nmat.book.....S,2016ApJ...826..200S,2020MNRAS.492.1385K} to update the turbulent kinetic energy $K_T$ together with the hydrodynamic Euler equation
\begin{equation}
    \pdd{}{t}K_{\rm T} + \pdd{}{x_j} \left( K_{\rm T} \widetilde{v_j} \right) + P_{\rm T} \pdd{\widetilde{v_j}}{x_j} = C_{\rm T} - D_{\rm T}.
\end{equation}
The terms on the left-hand-side describe the time evolution, the advection with the flow and the contraction, $C_{\mathrm{T}}$ and $D_{\mathrm{T}}$ are called production and dissipation terms, $\widetilde{v_j}$ is the large-scale velocity and the turbulent pressure is given by $P_\mathrm{T}=2/3 K_\mathrm{T}$.

This formalism allows us to describe the advection, contraction and dilution of turbulent kinetic energy, together with a source term modelling the injection of turbulence by gas shearing motion as well as a decay term modelling turbulence dissipation.

Once we know the turbulent velocity dispersion in each cell we can describe the gas density distribution in this cell through a log-normal probability distribution function (PDF)
\begin{equation}
p(s)=\frac{1}{\sqrt{2\pi \sigma_s^2}}\exp {\frac{(s-\overline{s})^2}{2\sigma_s^2}},
\label{eq:pdf-distr} 
\end{equation}
with the logarithmic density $s=\ln (\rho / \overline{\rho})$, where $\rho$ is the local density and $\overline{\rho}$ the mean density of the cell.
The mean logarithmic density $\overline{s}=-1/2 \sigma_s^2$ is related to the standard deviation by
\begin{equation}
\sigma_s^2 = \ln \left(1 + b^2 \mathcal{M}^2\right),
\end{equation}
where we have introduced the Mach number defined by $\mathcal{M} = \sigma_T/c_s$ and where $b$ is the turbulence forcing parameter (solenoidal or compressive). We adopt a value of $b=0.4$, corresponding to mixed turbulent forcing \citep[][]{2010A&A...512A..81F,2012ApJ...761..156F}. It is clear from this formalism that a larger turbulent velocity dispersion will result in a wider density PDF.

We can use the standard energy spectrum for isothermal supersonic turbulence
$\sigma(\ell) = \sigma_T \left( \ell / \Delta x \right)^{1/2}$,
to prolong the turbulent velocity dispersion to any smaller unresolved scales $\ell<\Delta x$.
A critical unresolved scale is the sonic length, defined by 
\begin{equation}
\sigma(\ell_s) = c_s~~~\mathrm{so~that}~~~\ell_s = \Delta x / \mathcal{M}^2.
\label{eq:sonic}
\end{equation}
Below the sonic scale, density fluctuations become very weak. Clouds at and below this scale can be viewed as quasi-homogeneous density regions. This simple model is used for both our H$_2$/CO model and our star-formation recipe.

\subsection{Subgrid model for \texorpdfstring{H$_2$}{H2} and CO}

We use a subgrid model similar to \cite{2018MNRAS.473..271V}, with relatively minor differences presented in more details in a companion paper (Kretschmer \textit{et al.}, in prep). We nevertheless summarise here briefly its main characteristics. We consider that our computational cells are divided into tiny volume elements of size equal to the sonic length. The homogeneous density in these subsonic volume elements is following the log-normal PDF that we have introduced earlier. 

The formation of H$_2$ and CO follows the simple approximate model of \cite{2012MNRAS.421..116G}. A key ingredient of this equilibrium chemical model is the radiative transfer of far-UV radiation. For this, we use a simple approximation where the column density of absorbing material (dust, HI or H$_2$) is obtained using the local Jeans length of the subgrid volume elements, assuming a fixed temperature of 10~K. As demonstrated by \cite{2017MNRAS.465..885S}, this local Jeans length approximation is surprisingly accurate, when compared to more realistic radiative transfer approaches.
We also take into account the effect of self-shielding from overlapping H$_2$ lines using the formula provided by \cite{1996ApJ...468..269D}. 

The input interstellar radiation field is approximated for the entire galaxy using $G=G_0 (\mathrm{SFR / M}_\odot \mathrm{yr}^{-1})$, where SFR is the global galaxy star formation rate and $G_0$ is the classical Habing flux in the solar neighbourhood. The CO line emission (focusing here only on the 1--0 transition) is computed using a standard escape probability method and assuming local thermodynamical equilibrium (LTE). The CO line width is modelled self-consistently using our subgrid velocity dispersion for turbulence. This last component of the model turns out to be crucial for setting the CO luminosity of the molecular gas, as turbulent line broadening decreases the optical depth, allowing more photons to escape.

\subsection{Subgrid model for star formation}

Star formation in our simulations is modelled using a standard Schmidt law, where we define the star formation density as $\dot{\rho}_\star=\epsilon_\mathrm{ff} {\rho}/{t_\mathrm{ff}}$,
where $\rho$ is the density of the gas, $t_\mathrm{ff}=\sqrt{{3\pi}/{(32 G \rho)}}$ is its free-fall time and $\epsilon_\mathrm{ff}$ is the star formation efficiency per free-fall time. Compared to more classical recipes where $\epsilon_\mathrm{ff}$ is a constant, our approach differs in the sense that $\epsilon_\mathrm{ff}$ is allowed to vary from cell to cell, based on the subgrid turbulent state of the gas.

The formalism we use to compute the local star formation efficiency is similar to the one we use for molecular chemistry.
We first reconstruct the density PDF (Eq.~\ref{eq:pdf-distr}) in each cell based on the turbulent Mach number $\mathcal{M}$.
We then compute a critical density above which the gas in the subgrid volume elements will collapse and form stars.
This critical density is obtained by requiring that the virial parameter of the subgrid volume element is less than one, 
which is equivalent to requiring the sonic length (Eq.~\ref{eq:sonic}) to be smaller than the local Jeans length
\citep{2005ApJ...630..250K,2006A&A...450..265S,2012ApJ...761..156F,2016ApJ...826..200S}.

The resulting efficiency per free-fall time can vary widely between $0\%$ and $100\%$, depending on the local conditions in each cell. 
Such a model is capable of producing \textit{at the global scale of the entire galaxy} very high star formation efficiencies $\simeq 20\%$ in starbursts, 
as well as very low global efficiencies $\simeq 0.1\%$ in quenched galaxies. In normal star-forming galaxies such as the one discussed in this paper, 
the predicted global efficiency at low redshift is usually of the order of $1\%$ \citep{2015IAUGA..2257403P,2016ApJ...826..200S,2017MNRAS.470..224T,2018MNRAS.478.5607T,2021A&A...651A.109D,2020MNRAS.492.1385K, 2020MNRAS.497.4346K,2021MNRAS.501...62N}.

\subsection{Additional galaxy formation recipe}

SN explosions inject thermal energy of $E_\mathrm{SN}=10^{51} \mathrm{erg}$ into the surrounding gas.
If the cooling radius of the corresponding Sedov blast wave is not resolved by the grid, which usually occurs at high gas densities, we inject also the proper amount of momentum into the surrounding ISM \citep{2015MNRAS.450..504M}. SN explosions are resolved in time, in the sense that a star particle will feature multiple individual SNe between $3$ Myr and $20$ Myr after its birth.
We finally briefly recap the remaining ingredients for our galaxy formation model: metal enrichment is modelled using a yield of $y=0.1$ per exploding massive star.
Cooling and heating of the gas is modelled using standard H and He cooling processes. We also include metal cooling, heating by a standard UV cosmic background and self-shielding of gas \cite[see, e.g.][for more details]{2013MNRAS.429.3068T,2016MNRAS.457.1722R,2017MNRAS.471.2674R,2017MNRAS.469..295B,2018MNRAS.475.5688B}.

\section{Results}
\label{section:results}
We now present the results for our fiducial simulated galaxy.
We use this simulation as a numerical experiment to explore the possibility to obtain cold, rotation-dominated galaxies at high redshift within the standard $\Lambda$CDM model of galaxy formation. Our analysis focuses on exploring one specific example of a very high resolution simulation (to be followed soon with a statistical sample in a follow-up paper).
\begin{figure*}
    \centering
    \includegraphics[width=\textwidth]{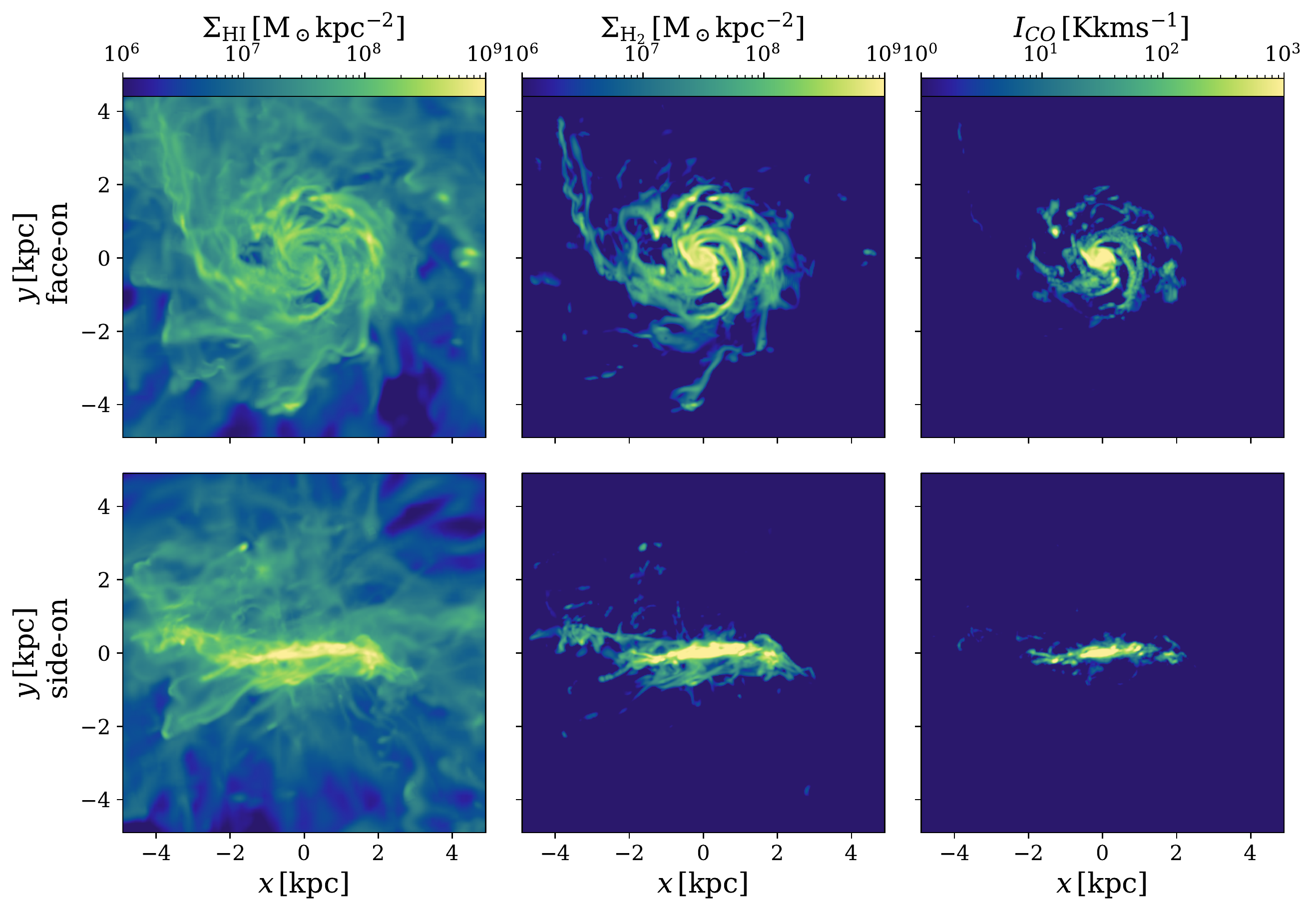}
    \caption{Maps of the simulated galaxy at $z=3.5$ from left to right: surface density of HI, surface density of H$_2$ and CO intensity map. Top-row shows the galaxy viewed face-on and the bottom-row shows the galaxy viewed edge-on. In comparison to HI, we see that H$_2$ and CO trace denser and less turbulent gas that is closer to the mid-plane of the disc.}
    \label{fig:maps}
\end{figure*}
\begin{figure*}
    \centering
    \includegraphics[width=\textwidth]{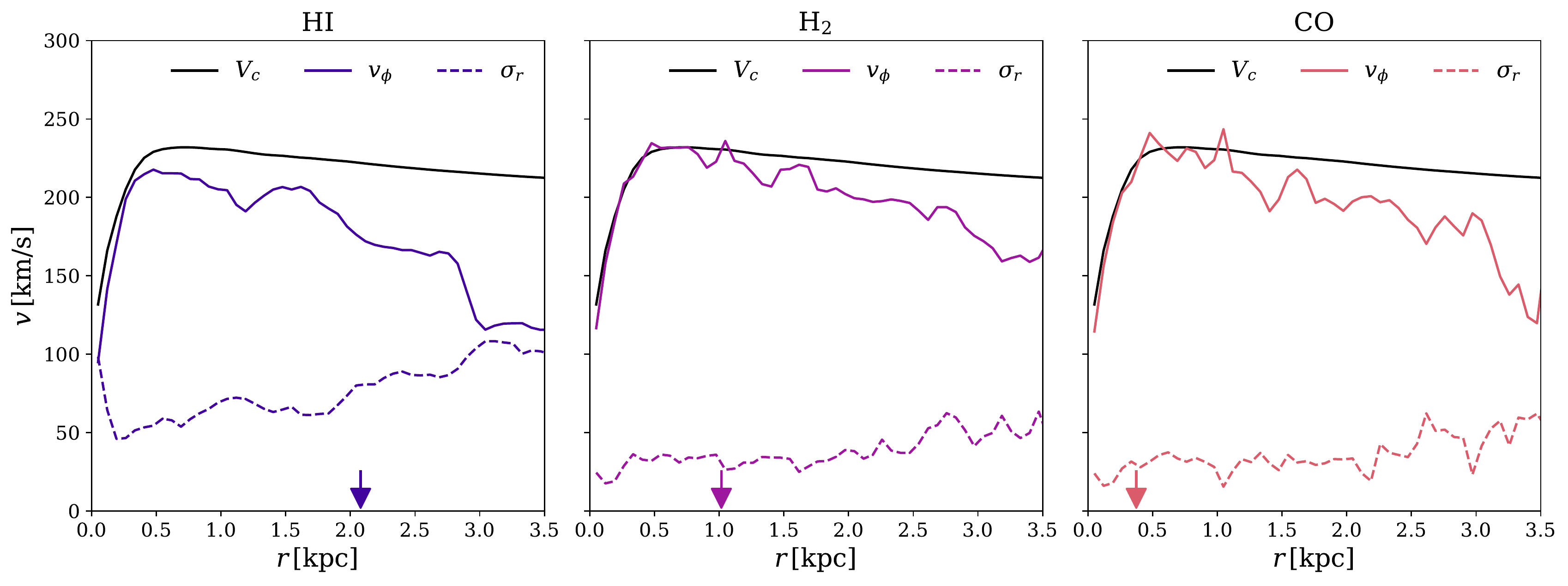}
    \caption{Velocities measured at $z=3.5$ using the three different tracers HI, H$_2$ and CO. Shown is the circular velocity $V_c$ together with the rotational velocity $v_\phi$ and the radial velocity dispersion $\sigma_r$. We see that the obtained $v_\phi$ from H$_2$ and CO is larger than the one obtained from HI and closer to $V_c$. Furthermore, the dispersion is smaller and remains mostly constant with a value around $\sigma_r \sim 30  \, \mathrm{km \, s^{-1}}$ for H$_2$ and CO. For HI, the dispersion increases with radius at $r>2$~kpc, where there is little H$_2$, and no CO. The relative difference in the obtained values for $\sigma_r$ is larger than those for $v_\phi$. The arrows mark the half-mass radius $r_e$ for each component.}
    \label{fig:kinematics}
\end{figure*}
\begin{figure}
    \centering
    \includegraphics[width=\columnwidth]{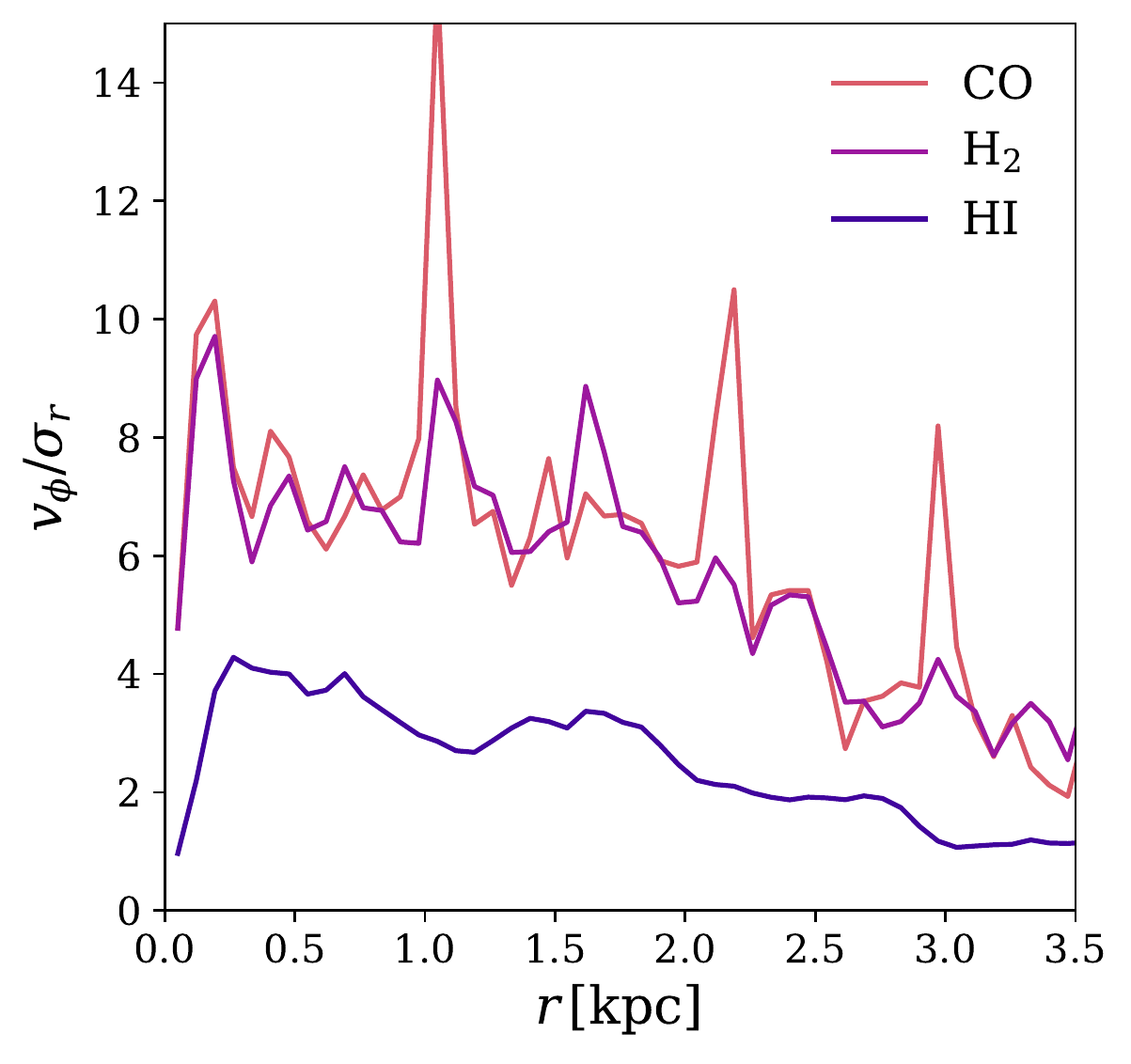}
    \caption{The resulting ratio of rotational velocity to velocity dispersion in the simulated galaxy at $z=3.5$ obtained from different gas tracers. Large values around $\simeq 8$ are obtained inside $2$ kpc using CO and H$_2$ because they probe denser and less turbulent gas close to the mid-plane of the disc. For certain radii, $\vsig>10$ for CO. Using HI we obtain values around $\simeq 3-4$ for $r<2$ kpc.}
    \label{fig:v_sig}
\end{figure}
\begin{figure*}
    \centering
    \includegraphics[width=\textwidth]{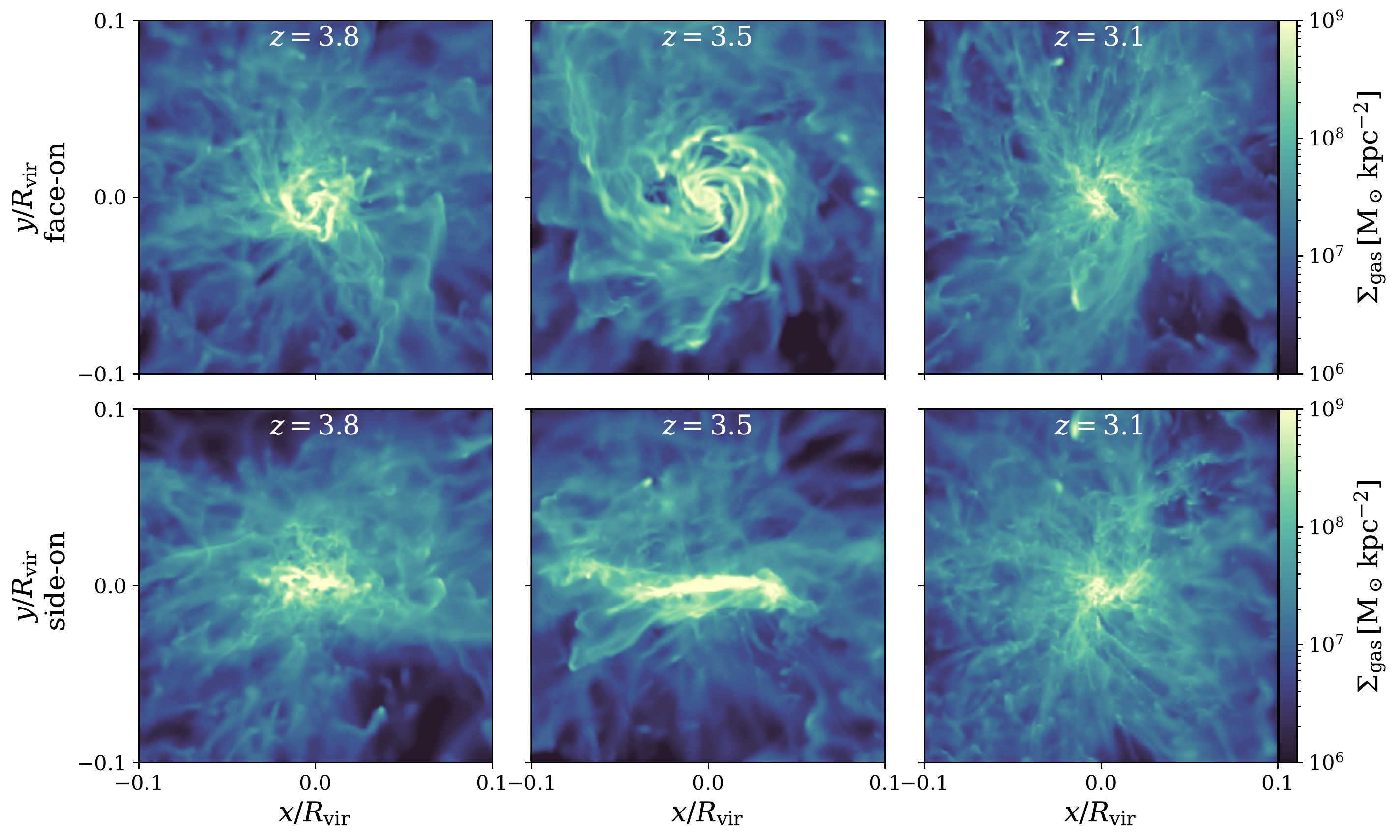}
    \caption{The surface density of our galaxy at three redshifts evolving from left to right with time, using all the gas. Top-row shows the face-on projection and bottom row the edge-on. The middle column at $z=3.5$ represents the epoch of peak $\vsig$. The duration of this disc phase is $\sim 410$ Myr, as discussed in the text. It is apparent that before and after this epoch, the gas disc is very compact with no clear visible extended disc structure.}
    \label{fig:pre_post_tgt}
\end{figure*}
\subsection{Kinematic analysis}
We identify DM halos on-the-fly using the clump finder \texttt{PHEW} implemented in \texttt{RAMSES} 
\citep{2015ComAC...2....5B, 2022MNRAS.510..959I}.
\texttt{PHEW} uses an over-density criterion to identify density peaks and provides halo masses as well as positions of the density peaks.
In post-processing, we use the masses and positions obtained by the halo finder as a first guess. We then use the iterative shrinking sphere algorithm using the DM particles to obtain the true halo centre, virial radius and virial mass following the definition of \cite{1998ApJ...495...80B}.
The galaxy centre is obtained using again the shrinking sphere algorithm, although this time it is applied to the stellar particles, using the previous halo centre as an initial guess.
We define the galaxy as the entire region within a sphere of radius $0.1 R_\mathrm{vir}$. Finally, we define the bulk velocity of the system as the average velocity of stars and DM particles inside a sphere of radius 1~kpc.

We first analyse our simulated galaxy at $z=3.5$, where it features a rotation-dominated gas disc. We will study later its time evolution.
The halo mass at this redshift is $M_\mathrm{vir} = 3.7 \times 10^{11} \mathrm{M_\odot}$ and the viral radius is $R_\mathrm{vir}=51$ kpc.
We find that the stellar half-mass radius is $r_{e,\star}=0.5$ kpc and the gas half-mass radius is $r_\mathrm{e, gas}=1.6$ kpc. The radius of the disc is $r_\mathrm{d}=2.6$ kpc\footnote{Following \cite{2014MNRAS.443.3675M}, $r_\mathrm{d}$ contains 85\% of the cold mass within a cylinder of radius $0.1 R_\mathrm{vir}$ and height $1$ kpc.}.
The stellar mass is $M_\star=1.0 \times 10^{10} \mathrm{M_\odot}$, where most of the stars are located in a massive bulge with mass $M_{\star,b}=7.3 \times 10^{9} \mathrm{M_\odot}$. The gas fraction is $f_\mathrm{gas}=M_\mathrm{gas}/(M_\star + M_\mathrm{gas})=0.28$, smaller than the average $f_\mathrm{gas}\sim 0.5$ observed in high-redshift star-forming galaxies \citep[][]{2020ARA&A..58..157T}.

We now want to study the distribution of different gas species.
We show in \autoref{fig:maps} face-on and edge-on maps of our galaxy at $z=3.5$, oriented relative to the angular momentum of the gas. 
Shown are the surface density of HI and H$_2$, as well as the CO $1-0$ line intensity.
It is apparent that HI is more extended and covers a larger surface area than H$_2$ and CO.
Indeed, we see that H$_2$ and CO are concentrated close to the mid-plane of a dense and thin gas disc. 
Most of the H$_2$ mass is concentrated in a few massive clumps, in the spiral arms and in the bulge, 
three regions where large gas densities are expected.
We note that H$_2$ seems to be slightly more extended than the CO line emission. The more diffuse H$_2$ component is associated to freshly accreted, lower-density material that is mostly CO dark.

We now turn to the kinematic analysis of our galaxy. We are interested in understanding how the measured rotation velocity $v_\phi$ to velocity dispersion $\sigma_r$ ratio $\vsig$ changes using our different gas tracers.
Therefore, we analyse separately the kinematics for the gas traced by HI, H$_2$, and CO.
We first calculate the angular momentum using all the gas inside $0.05 R_\mathrm{vir}$ and orientate the disc relative to it. We have also performed the orientation for each component separately, which did not change our results significantly.
The tangential and radial velocities $v_\phi$ and $v_r$ are calculated in the plane of the disc.
Mass-weighted average values are computed in cylindrical rings with width $\Delta r=60$~pc and height $h=4$~kpc\footnote{{This is motivated by the fact that observations integrate all the gas along the line-of-sight. Note that $\sim87\%$ of the total gas mass is within $|z|<1$~kpc.}}.
The radial velocity dispersion is calculated as 
\begin{equation}
\sigma_r = \left(\langle v_r^2\rangle-\langle v_r\rangle^2\right)^{1/2},
\end{equation}
where the average values are computed in cylindrical rings as well.
The circular velocity $V_c$ is calculated with the total mass $M$ enclosed in a sphere of radius $R$ as
$V_c^2 = GM(<R)/R$,
where $G$ is the gravitational constant.

\autoref{fig:kinematics} shows the velocity profiles measured in the simulation at $z=3.5$.
It is apparent that the tangential velocity $v_\phi$ measured using H$_2$ and CO is closer to the circular velocity $V_c$ than the one obtained using HI. We also see that the radial velocity dispersion $\sigma_r$ measured using HI is larger than the one measured using H$_2$ and CO by a factor of $2.1$ and $2.3$, respectively. Furthermore, the kinematic analysis based on HI shows a strongly declining rotation curve and a steeply increasing velocity dispersion with radius from $50\,\mathrm{km\,s^{-1}}$ up to $100\,\mathrm{km\,s^{-1}}$ whereas the kinematic analysis based on H$_2$ or CO features only a mildly decreasing rotation curve and a velocity dispersion that is almost flat around $\simeq 30 \, \mathrm{km \, s^{-1}}$.

The discrepancy between $V_c$ and $v_\phi$ at large radii is explained by the increasing pressure support \citep[][]{2014A&A...566A..71L,2020MNRAS.497.4051W,2021MNRAS.503.5238K}. We note that the purpose of our kinematic analysis is to obtain values for $v_\phi$ and $\sigma_r$. A precise study, relating the observable velocities to the exact velocity profile given by the underlying mass-distribution is beyond the scope of this paper \citep[this has been studied e.g. in][]{2021MNRAS.503.5238K}

Combining these different kinematic measurements, we obtain the radial profiles of $\vsig$ that we show in \autoref{fig:v_sig}.
We see that the values for $\vsig$ obtained using H$_2$ and CO are larger than those obtained using HI on average by factors of $2.4$ and $2.6$, respectively. At the half-mass radius of each component, $\vsig$ are $8.1,9.0,2.2$ for CO, H$_2$ and HI respectively and at a fixed radius of $r=1.5$ kpc, $\vsig$ are $7.6,6.4,3.2$ for CO, H$_2$ and HI respectively.
It is also apparent that using the molecular tracers, $\vsig$ reaches values as large as $\simeq 8$ within $r_\mathrm{e,gas}$, whereas using HI yields values as low as $\simeq 3$. Furthermore, we see that in some rings $\vsig>10$ for CO. We emphasise that the differences between the values for $\vsig$ obtained using the different tracers originates mostly from the different values we obtained for $\sigma_r$.
This implies that using a tracer that probes predominantly dense gas increases the rotational to dispersion ratio by a factor of $\gtrsim2.4$.

\subsection{Kinematics during disc assembly}
In order to estimate how frequently such a large $\vsig$ ratio would be detected in observations, we now examine the time evolution of our galaxy.
We focus on average values and follow the commonly adopted definition of $v_\mathrm{max}/\langle \sigma \rangle$, where $v_\mathrm{max}$ is the maximum of $v_\phi$ and $\langle \sigma \rangle$ is the average radial velocity dispersion, both inside $2$~kpc \citep[see, e.g.][]{2020Natur.584..201R}.
We include measurements done using all the gas from our simulations. This is usually not directly available in observations but allows for further insights when comparing to the individual tracers.

\begin{figure}
    \centering
    \includegraphics[width=\columnwidth]{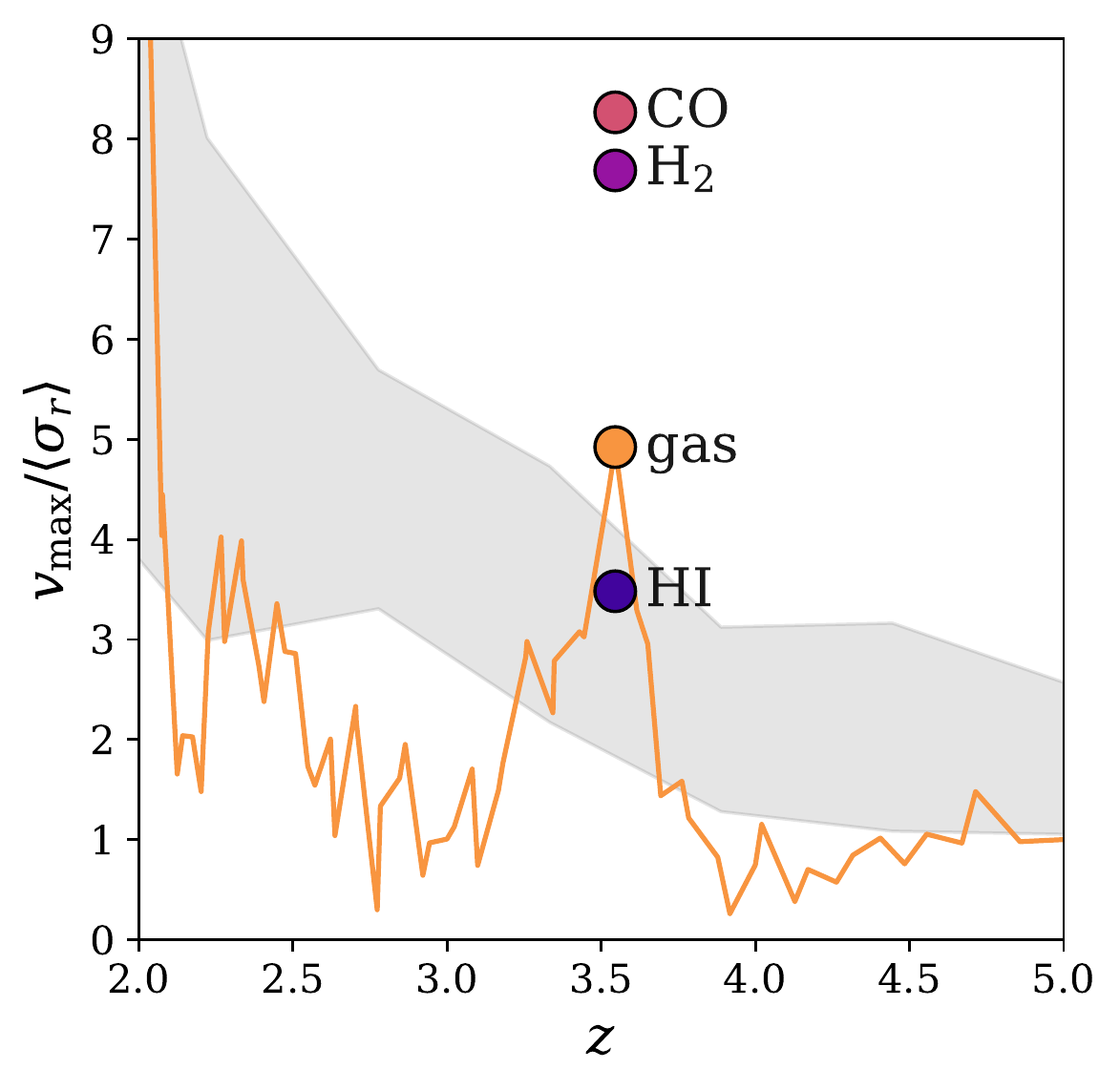}
    \caption{
    The evolution of the ratio of maximum rotation velocity to mean velocity dispersion (inside 2~kpc) $\vmaxsig$ as a function of redshift.
    Shown as orange curve is the evolution of $\vmaxsig$, measured using all the gas in our simulated galaxy.
    The circles show the values obtained using the different tracers at $z=3.5$.
    The grey band shows the $1\sigma$ confidence interval of $\vmaxsig$ as a function of time, obtained from a set of galaxies.
    }
    \label{fig:v_sig_vs_z}
\end{figure}

In \autoref{fig:v_sig_vs_z}, we show the evolution of $\vmaxsig$ as a function of redshift. We see that, using all the gas, $\vmaxsig\simeq 1$ at high redshift but then very rapidly rises up to $\vmaxsig\simeq 5$ at $z=3.5$.
This rapid rise is related to an epoch of efficient gas accretion, where cold streams transport pristine gas to the disc and add angular momentum in a constructive way. Namely, the accreted gas is co-rotating and mostly co-planar \citep[][]{2012MNRAS.422.1732D,2015MNRAS.449.2087D}.
We have already identified in earlier work a similar event around $z\sim1$, and argued that co-planar and co-rotating gas accretion could be responsible for the fast assembly of the MW disc. We called this event the Grand Twirl \citep{2020MNRAS.497.4346K}.

Our present simulation features a high-redshift analogue of the Grand Twirl.
Moreover, at this epoch, the virial mass of the halo is already above the critical mass
$M_\mathrm{vir}\sim 2 \times 10^{11}\mathrm M_\odot$ where discs are prone to survive merger-driven spin flips \citep[][]{2020MNRAS.493.4126D}.
We indeed find that the direction of the angular momentum remains almost constant during the period where the cold disc is present, namely from $z=3.9$ to $z=3.2$, which corresponds to a time of $t\simeq 410$ Myr.
Using for the disc orbital time $t_\mathrm{orb} = 2\pi r_d/v_\mathrm{max} \simeq 80$~Myr, we conclude that the cold disc survives for roughly 5 orbits.
Furthermore, expressed in units of the virial time $t_\mathrm{vir} = R_\mathrm{vir}/V_\mathrm{vir} \simeq 260$ Myr at $z=3.6$, the duration of the cold disc phase is $1.6 t_\mathrm{vir}$ \citep[][]{2013MNRAS.435..999D}. The cold gas disc survives for a particularly long period of time.

After $z=3.2$, $\vmaxsig$ decreases to $1$ again, because the gas disc is destroyed by a series of unfortunate events, namely counter-rotating and out-of-plane accretion of gas and satellites, until it rapidly rises back again to $10$ at $z\sim 2$.
From the surface density plots in \autoref{fig:pre_post_tgt} we can see this dramatic evolution of the gas disc before, during and after the above described short epoch of transient disc assembly.
It is apparent that at $z=3.8$, before the cold disc phase, the gas is in a very compact and thick disc. At $z=3.5$, during the peak $\vsig$ epoch, an extended and thin disc is formed and after this epoch, at $z=3.2$, no disc-like structure is visible anymore.
Quantitatively, this is confirmed by measuring the half mass radii of the gas which gives $r_e = (1.13, 1.6, 2.4)$~kpc, and the rotational velocities which are $v_\phi = (138.7, 210.1,  41.7)\,\mathrm{km\,s^{-1}}$ at $z=3.8, 3.5, 3.1$, respectively, resulting in specific angular momenta of $j_\mathrm{gas}=(157.2, 342.2, 102.0)\,\mathrm{kpc\,km\,s^{-1}}$.
Again, these values demonstrate the ephemeral nature of the formation of cold, rotation-dominated gas disc at high redshift.

In \autoref{fig:v_sig_vs_z}, we also compare the rather erratic history of our galaxy with the typical $\vmaxsig$ of the MIGA sample, where we have computed the evolution of $\vmaxsig$ using all the gas for each galaxy following the same methodology as described above. Finally, we compute in redshift bins the median values and show the resulting $1\sigma$ confidence interval in the figure.
We see that $\vmaxsig$ of our galaxy first rises significantly above and then drops below the typical evolution of the general galaxy population.

Finally, we also show in \autoref{fig:v_sig_vs_z} as coloured circles the values for $\vmaxsig$ obtained using the different tracers and obtained using all the gas at $z=3.5$. 
As discussed above, using a molecular tracer increases $\vmaxsig$  by a factor of $2.4$ compared to HI ($\vmaxsig = 3.4$), such that we find $\vmaxsig = 8.3$ using CO and $\vmaxsig = 7.7$ using H$_2$. We also see that HI gives slightly smaller values than using all the gas because it traces more extended and more diffuse gas.

In summary, we have seen that the large value of $\vmaxsig$ obtained using CO and H$_2$ in our simulation is a result of two effects.
First, it coincides with the climax of the assembly history of a particularly cold disc. Second, the measured value of $\vmaxsig$ is enhanced by tracing specifically the cold gas close to the mid-plane.
\section{Discussion}
\label{section:discussion}
In the high-redshift Universe, gas discs are on average not expected to be highly rotation dominated, which is confirmed by simulations and indicated by ionized gas observations \citep[][]{1996ApJ...462L..17S,1999ApJ...519....1S,2004ApJ...604L..21E,2006Natur.442..786G,2018MNRAS.473.1930E,2019MNRAS.490.3196P}.
Streams of cold gas feed the galaxy with pristine gas from the cosmic web maintaining a large gas fraction, which enable fragmentation of the gas discs into massive clumps and clumpy rings \citep[][]{2005MNRAS.363....2K,2007ApJ...670..237B,2008MNRAS.390.1326O,2009Natur.457..451D,2009ApJ...703..785D,2011ApJ...741L..33B,2015ApJ...814..131G,2016ApJ...827...28G,2020MNRAS.496.5372D}.
Clearly, these violent disc instabilities may disturb the integrity of the disc.
Nevertheless, the most destructive mechanism that is common at high redshift are mergers, especially major mergers.
It is therefore only in massive galaxies with $M_\mathrm{vir}\gtrsim 2 \times 10^{11}\mathrm M_\odot$ where discs are expected to survive because the galaxy merger rate is low compared to the orbital time \citep[][]{2020MNRAS.496.5372D}.  
This enables massive galaxies to be rotation-dominated, but typically not as cold as $\vsig\sim 10$ as found in recent observations \citep[][]{2020Natur.584..201R,2021A&A...647A.194F,2020Natur.581..269N,2021Sci...371..713L}.

An important mechanism to generate an extended gas disc that is rotation dominated at $z=0$ is a final episode of constructive gas accretion. Where by constructive we describe material entering the inner parts of the halo from streams that is co-planar and co-rotating. To build the disc of the MW, such an event \citep[referred to as the Grand Twirl in][]{2020MNRAS.497.4346K} probably occurred at $z\sim1$ to have sufficient material to build the disc and at the same time to reduce the likelihood of destructive events afterwards \citep[][]{2012MNRAS.423.1544S,2012MNRAS.422.1732D,2015MNRAS.449.2087D,2013ApJ...769...74S,2020MNRAS.497.4346K}.

We have seen by analysing our simulation, that such an episode of disc build-up is also possible at high redshift during a rapid disc assembly  episode. The main difference with the Grand Twirl is that this high-redshift analogue is only a transient phenomenon, since subsequent events quickly destroy the cold disc. However, this demonstrates that there is the possibility for a high-redshift cold and rotation-dominated disc as a result of such a mechanism.
As we have seen above, this is an episode that lasts $\sim 5$ orbital times.
By comparing this to the average time between mergers we see how particularly quiet the evolution of our galaxy is during this episode.
Namely, using the virial mass of our galaxy, the time between mergers is $\sim 2$ orbital times \cite[see Eq. 15 of][]{2020MNRAS.493.4126D}. The stellar mass of our galaxy implies for a typical galaxy a virial mass of $\sim 10^{12} \mathrm{M_\odot}$ \citep[e.g.][]{2019MNRAS.488.3143B} that results in a slightly longer time between mergers of $\sim 3$ orbital times.

This episode could be observed by analysing the gas kinematics through specific gas tracers.
Especially telescopes like ALMA enable detailed kinematic studies using cold-gas tracers like CII or CO.
HI maps of individual galaxies can be obtained with SKA, although probably not at redshifts as high as $z\sim 3.5$.
As we have seen above, different tracers follow different gas components, meaning that for example a molecular tracer will follow cold and dense gas, whereas HI will track more diffuse gas. The inferred kinematics are therefore different.

Since cold streams at high redshift are common phenomena, epochs of rapid disc growth are expected if the accreted material is joining the disc co-rotating and co-planar.
We speculate that many massive galaxies at high redshift undergo such constructive events that produce large $\vmaxsig$ for a short epoch in their formation history.
Disc-instabilities, violent feedback and mergers will destroy those discs, explaining why large $\vsig$ at high redshift are only achieved during brief episodes in a galaxies history. 
Only for galaxies above $M_\mathrm{vir} \sim 2 \times 10^{11} \mathrm{M_\odot}$, which are more rare at high redshift, mergers are less and less frequent compared to the orbital time $t_\mathrm{merger} \propto (M_\mathrm{vir}/10^{11} \mathrm{M_\odot})^{0.62} t_\mathrm{orb}$ \citep[Eq. 15 of ][]{2020MNRAS.493.4126D}, such that discs in those halos survive longer with respect to the orbital time. Consequently, the increased number of massive galaxies in the present universe together with the reduced likelihood of destructive events, allows for long-lived gas discs with larger $\vsig$ growing undisturbed.

Additionally to these effects that can cause large $\vsig$, we expect a host-mass dependence.
This is in particular interesting for high-redshift observations of galaxies, since many focus on massive systems. Observations as well as simulations tend to indicate that the dispersion stays approximately constant and therefore the $\vsig$ ratio is expected to very crudely increase with the galaxy mass, as for example visible in Fig. 2 of \cite{2020MNRAS.496.5372D}.

We note that simulations use specific sets of different subgrid models which could influence the emergence of gas discs.
For example, the details of a particular feedback model could promote or destroy discs by using a comparably weaker or stronger implementation.
However, we have seen that our particular set of subgrid models available in \texttt{RAMSES} predicts galactic properties that are in good agreement with observations.
A weaker stellar feedback would likely promote the formation of rotation-dominated gas discs, but at the same time result in the wrong SFR histories, stellar masses and baryon fractions at $z=0$. As a consequence, constraints on the $M_\star/M_\mathrm{vir}$ relation obtained using the abundance matching technique would be violated.
Nevertheless, different implementations, combination of subgrid models, or models that include additional processes that we do not consider, like an AGN, may also reproduce observed quantities together with smaller $\vsig$. Studying in detail the effect of these different subgrid models is however beyond the scope of this paper.

An important aspect of this paper is the effect of using different tracers on the inferred kinematics. We did restrict ourselves to CO, H$_2$ and HI but in principle other tracers can be build into the model. One interesting extension would be the study of ionised gas. For example H$\alpha$ and OIII are regularly used to observed galaxies at $z=1-3$ and those observations provide crucial evidence for turbulent discs.
Furthermore, the line of singly ionised carbon CII, would be a particularly interesting extension to our model since it is often used to identify discs with very large $\vsig$ at high redshift. A caveat might be that CII emission arises from both atomic and molecular phases \citep[e.g.][]{2021ApJ...915...92T}. Depending on the fraction originating from the two phases, the corresponding emission in our model would be in between the individual HI and H$_2$ emissions.

One particularly important questions remains unanswered. Episodes of co-rotating and co-planar gas accretion events are expected to be common at high redshift. It is unclear, how often they lead to very cold discs and how long on average they survive. Therefore, our study of one individual galaxy should be followed up by a statistical sample. However, even if we could perform a large-scale simulation with similar resolution to the one presented here, we would need to mock-observe the galaxies in the simulation to compare it to current observations like ALMA. Different observational effects would probably influence the obtained values for $\vsig$.

\section{Conclusions}
\label{section:conclusions}
Recent observations suggested the existence of cold gas discs with $\vsig \sim 10$ at high redshift. Although rotation-supported discs are expected at high redshift, these findings of extremely cold gas discs are surprising.
Using a high-resolution zoom-in cosmological simulation, we have identified a period of an emerging cold gas disc at high redshift.
Our analysis revealed two interesting kinematic properties of the gas disc in our simulated high-redshift galaxy.
\begin{itemize}
    \item First, we have seen that this galaxy experiences an epoch when the rotation-to-dispersion ratio increases from $\vsig\simeq1$ to $\vsig\simeq5$ at $z=3.5$, namely $\vsig$ was temporarily increased by a factor of $\sim5$.
    The disc growth and survival is enabled by intense accretion of co-planar, co-rotating gas via cold cosmic-web streams into a cold disc.
    This epoch gives rise to values larger than the typical $\vsig$ from a sample of galaxies.
    Only after $\sim 5$ orbital times, the discs is destroyed by counter-rotating streams and mergers so that $\vsig$ decreases back to $\vsig\simeq 1$.
    \item Secondly, we have shown that using different tracers significantly influences the resulting value of $\vsig$. 
    For atomic HI gas, $\vsig \simeq 3.5$, but for molecular CO or H$_2$ the obtained values are $\simeq 8$.
    This reflects that molecular gas is mostly encapsulated in dense and cold molecular clouds, which reside close to the disc mid-plane, while the atomic gas is more spread out into a turbulent and extended thicker disc.
\end{itemize}

Finally, we note that from a theoretical point of view, episodes of co-rotating and co-planar gas accretion events are expected to be common at high redshift. However, it remains unclear how often they lead to very cold discs and how long on average they survive. To evaluate if the proposed mechanisms occurs frequently enough to explain the multiple discoveries of cold gas disks in high-z galaxies, proper statistical measurements in both simulations and observations are required.

\section*{Acknowledgements}
The authors thank the referee for their constructive comments that improved the quality of the paper.
We acknowledge stimulating discussions with Volker Springel, Pedro R. Capelo, Lucio Mayer, Robert Feldmann, Francesca Rizzo, Simona Vegetti and Xavier Prochaska.
This work was supported by the Swiss National Supercomputing Center (CSCS) project s1006 - ``Predictive models of galaxy formation'' and the Swiss National Science Foundation (SNSF) project 172535 - ``Multi-scale multi-physics models of galaxy formation''.
This work was also partly supported by the Israel Science Foundation grant ISF 861/20 and by the German-Israel DIP grant STE1869/2-1 GE625/17-1.
The simulations in this work were performed on Piz Daint at the Swiss Supercomputing Center (CSCS) in Lugano, and the analysis was performed with equipment maintained by the Service and Support for Science IT, University of Zurich.
We made use of the \texttt{pynbody} package \citep{2013ascl.soft05002P}.

%%%%%%%%%%%%%%%%%%%%%%%%%%%%%%%%%%%%%%%%%%%%%%%%%%
\section*{Data Availability}
The data underlying this article will be shared on reasonable request to the corresponding author.



\bibliographystyle{mnras}
\bibliography{co_h2}

%%%%%%%%%%%%%%%%%%%%%%%%%%%%%%%%%%%%%%%%%%%%%%%%%%

%%%%%%%%%%%%%%%%% APPENDICES %%%%%%%%%%%%%%%%%%%%%

% \appendix

% \section{Additional plots}


%%%%%%%%%%%%%%%%%%%%%%%%%%%%%%%%%%%%%%%%%%%%%%%%%%

% Don't change these lines
\bsp	% typesetting comment
\label{lastpage}
\end{document}